\documentclass[twocolumn]{aastex63}
\usepackage[normalem]{ulem}     

\newcommand{\p}[1]{{\color{red}{#1}}}

\newcommand{\Vp}{$V_\mathrm{p}$}
\newcommand{\Tp}{$T_\mathrm{p}$}
\newcommand{\np}{$n_\mathrm{p}$}

\received{?}
\revised{?}
\accepted{28 September 2021}

\submitjournal{ApJ}

\shorttitle{The two-step Forbush decrease by ICME substructures}
\shortauthors{Janvier et al.}
\graphicspath{{./}{figures/}}

\begin{document}

\title{The two-step Forbush decrease: a tale of two substructures modulating galactic cosmic rays within coronal mass ejections.}

\correspondingauthor{Miho Janvier}
\email{miho.janvier@universite-paris-saclay.fr}

\author[0000-0002-6203-5239]{Miho Janvier}
\affiliation{Universit\'e  Paris-Saclay,  CNRS,  Institut d’Astrophysique Spatiale, 91405 Orsay, France}
\affiliation{Laboratoire Cogitamus, 1 3/4 rue Descartes, 75005 Paris, France}

\author[0000-0001-8215-6532]{Pascal D\'emoulin}
\affiliation{LESIA, Observatoire de Paris, Universit\'e PSL, CNRS, Sorbonne Universit\'e, Universit\'e de Paris, 5 place Jules Janssen, 92195 Meudon, France}
\affiliation{Laboratoire Cogitamus, 1 3/4 rue Descartes, 75005 Paris, France}

\author[0000-0002-8707-076X]{Jingnan Guo}
\affiliation{School of Earth and Space Sciences, Science and Technology of China, Hefei, China}

\author[0000-0002-7680-4721]{Sergio Dasso}
\affiliation{CONICET, Universidad de Buenos Aires, Instituto de Astronom\'\i a y F\'\i sica del Espacio, CC. 67, Suc. 28, 1428 Buenos Aires, Argentina}
\affiliation{Universidad de Buenos Aires, Facultad de Ciencias Exactas y Naturales, Departamento de Ciencias de la Atm\'osfera y los Oc\'eanos and Departamento de F\'\i sica, 1428 Buenos Aires, Argentina}

\author[0000-0002-4017-8415]{Florian Regnault}
\affiliation{Universit\'e  Paris-Saclay,  CNRS,  Institut d’Astrophysique Spatiale, 91405 Orsay, France}

\author{Sofia Topsi-Moutesidou}
\affiliation{Erasmus+ program, University of Ioannina, Greece}

\author{Christian Gutierrez}
\affiliation{Universidad de Buenos Aires, Facultad de Ciencias Exactas y Naturales, Departamento de Ciencias de la Atm\'osfera y los Oc\'eanos and Departamento de F\'\i sica, 1428 Buenos Aires, Argentina}

\author[0000-0002-2137-2896]{Barbara Perri}
\affiliation{Centre for Mathematical Plasma Astrophysics, KU Leuven, Celestijnenlaan 200b-box 2400, B-3001 Leuven, Belgium}

\begin{abstract}
Interplanetary Coronal Mass Ejections (ICMEs) are known to modify the structure of the solar wind as well as interact with the space environment of planetary systems. Their large magnetic structures have been shown to interact with galactic cosmic rays, leading to the Forbush decrease (FD) phenomenon. 
We revisit in the present article the 17 years of Advanced Composition Explorer spacecraft ICME detection along with two neutron monitors (McMurdo and Oulu) {with a superposed epoch analysis} to further analyze the role of the magnetic ejecta in driving FDs.  {We investigate in the following the role of the sheath and the magnetic ejecta in driving FDs, and} we further show that for ICMEs without a sheath, a magnetic ejecta only is able to drive {significant} FDs of comparable intensities. Furthermore, a comparison of samples with and without a sheath with similar speed profiles enable us to show that the magnetic field intensity, rather than its fluctuations, is the main driver for the FD. 
Finally, the recovery phase of the FD for isolated magnetic ejecta shows an anisotropy in the level of the GCRs. We relate this finding at 1 au to the gradient of the GCR flux found at different heliospheric distances from several interplanetary missions. 

\end{abstract}


\section{Introduction} \label{sec:intro}


Cosmic rays (CRs) are high energy particles that constantly bombard the planets of the solar system. Their origin can be out of the solar system, in which case they are referred to as galactic CRs (GCRs). These provide a radiation background that is slowly varying in time due to the changes throughout the solar cycle of the magnetic fields in the heliosphere.  Other sources of CRs are the Sun itself, via the generation of Solar Energetic Particles (SEP events), as well as transients in the solar system that can accelerate particles. 
Their variations in time following the solar cycle are well documented \citep[see the review of e.g.][and references therein]{Potgieter2013}.
CR research began in early 20th century with balloon measurements, and continues today with coordinated, world-wide coverage of neutron monitors \citep{Simpson2000}
{, and also with other types of detectors, such as muon telescopes \citep[e.g.,][]{Munakataetal2014} or water Cherenkov radiation detectors  \citep[e.g.,][]{Dassoetal2012}.}
CRs in the interplanetary medium are also routinely measured on board interplanetary probes \citep[such as the Ulysses mission, see][]{Simpson1992}, which provide measurements at different helio-distances as well as helio-latitudes.

   
While the monitoring of GCRs provides insights on the modulation of the solar cycle throughout the years, more rapid variations are measured due to the passage of solar wind transients.  These so-called Forbush decreases \citep[FDs,][]{Forbush1937} correspond to a rapid reduction in the intensity of GCRs, followed by a slow recovery typically lasting several days. They are associated with the arrival of a variety of solar wind disturbances: 
coronal mass ejections (CMEs, including here their sheath and their ejecta) or solar wind interaction regions, formed when a fast solar wind encounters and overtakes a slow solar wind such as corotating interaction regions {(CIRs)} are all found to be possible sources of these decreases.

Of primary interest here, are CMEs which find their origin in the Sun's corona. They are large ejection of plasma and magnetic field traveling in the heliosphere, where they can be detected by interplanetary probes \citep[see the reviews of, e.g.,][]{Zurbuchen2006, Kilpua2017}. When detected in situ, we refer to these structures as interplanetary CMEs (ICMEs). 


The main magnetic field structure which is detected in situ within ICMEs is a magnetic ejecta (ME). It is characterized by a magnetic field intensity larger, and smoother in time than that of the surrounding solar wind.  If plasma measurements are available, a low proton temperature and a low plasma $\beta$ is also associated with the ME. In some cases, but not all, a smooth rotation of the magnetic field has been inferred. {Such cases are called magnetic clouds} \cite[MCs,][]{Burlaga1981}. The ejecta acts as a piston-like driver that accumulates solar wind when its speed is significantly larger than that of the {ambient} solar wind. The accumulated solar wind then forms a sheath region ahead of the ejecta. 

Because of these two substructures, namely the sheath and the ME, {ICMEs} have been found to be responsible for a two-step FD \citep{Cane1994}. We make the clear distinction here that the main effect in driving the Forbush Decrease in ICMEs and co-rotating interaction regions 
is the presence of the sheath, rather than simply the discontinuity or the shock at its front. {Indeed, many observations show that the} FD gradually decreases and is not restricted to a small time interval after the shock. With a typical FD extending within the magnetic ejecta, it seems unlikely that the shock itself may be responsible for the FD {\citep[see e.g.][]{Cane2000, Forstner2020}. Furthermore, previous studies have pointed out the physical mechanisms within the sheath that could be at the origin of the FD \citep[e.g.][]{Masias2016}}.

The two-step FDs are not always found \citep{Jordan2011}; more generally, a review on the drive of FDs by ICMEs can be found in \cite{Cane2000}.
Among the main results, it was found that the plasma speed, the magnetic field strength and its field fluctuations are all well correlated with the intensity of the FD \citep[see e.g.][]{Kumar2014,Dumbovic2011}, with variations from one event to another.


While the turbulence in the sheath has been postulated to be one of the main culprit for the FD, a MC can also drive FDs due to the high magnetic field intensity due to the expansion of magnetic field that constitutes MCs. This was shown in several studies providing samples of FDs associated with MCs \citep{Cane1993, Cane1994, Lockwood1991}. 
In particular, \cite{Belov2015} studied the profiles of FDs in relation to that of the magnetic field by providing {a statistical study of the MCs temporal profiles along with that of FDs}. 
They found that the minimum of the GCR flux is found close to the MC center and not at its edges, especially in the case of slow MCs. Such studies show the importance of considering the effects of different substructures in driving the FD.


Another powerful statistical approach is the use of the so-called superposed epoch analysis (SEA) method, which was first introduced by \cite{Chree1913} by linking time series of sunspots variations at the Sun with fluctuations of the Earth's magnetism. The idea behind the method is to normalize time series of different events {(and with eventually different time duration)}, with \textit{a priori} similar characteristics, so as to obtain an average profile of the events. The simplest normalization is to adjust the start of all events to the same point in time, which we will refer to as a one-bound SEA. In the case of events with well defined frontiers, which is the case of ICMEs since they are generally defined with three times (the ICME start time, the ejecta start and the ejecta end), the time axis in the SEA can be adjusted for each substructure. More precisely, the starting and ending times of each substructure are set to the same normalized times (i.e. a generic {normalized} $t_{start}$ and $t_{end}$). In such a case, we will refer to this as a multiple-bounds SEA.

   
In the context of the FDs, several authors have shown the importance of the SEA method in linking characteristics of events such as ICMEs and FDs. For example, \cite{Kumar2014} used this technique to study the impact of ICMEs and CIRs {on FDs.}
For this, they crossed different catalogs of events, joining different signatures of ICMEs and CIRs, and used neutron monitors from Kiel and Calgary. They used a one-bounded superposed epoch, where the disturbance times are all shifted to a generic $t_{start}$. For ICMEs, they found that ICMEs associated with halo CMEs {launched in the direction of the observer}, MCs, and shocks are 1.5 to 4 times more effective in modulating GCRs than ICMEs not associated with these structures. The characteristic recovery time of GCR intensity due to the presence of these characteristics was also found to be longer than those due to ICMEs not associated with them.   

   
However, the limitation of the one-bound SEA is that because it is only bounded in one time \citep[ICME start time,
or the ME start time, as further studied in][]{Badruddin2016}, it is generally difficult to pinpoint the different substructure parts that may influence the FD. Another solution is to define other frontiers, as was done by \cite{Masias2016}. In their study, the authors performed a three-bounded SEA where the ICME start, the MC start and the MC end are used for the time normalization. Then, by ranking the ICMEs by order of mean MC speed (from slow, to medium and fast MC), they showed that the drop in the flux of GCR was mostly driven by the sheath in the case of slow MCs. Fast MCs, which also have a more turbulent sheath, and a stronger magnetic field intensity, were found to be associated with a two-step FD. 


With all the observations now available 
{to link FD events with ICMEs,} a remaining question is what the link between the magnetic intensity, the fluctuations, and the speed of ICMEs with the intensity and profiles of FDs is. The \textit{ForbMod} model proposed by \cite{Dumbovic2018, Dumbovic2020} is a step in this direction: this model combines the expansion and the cross-diffusion, and hence the FD happening inside MCs 
to explain the decay of the GCR flux during the passage of an ICME. 

Such a model can then be used to assess the MC expansion rate that explains the decrease in the intensity of FDs detected at different heliodistances, for example for the same events detected at different planets \citep[see the study of FDs detected at Earth and Mars,][]{Forstner2020}. Other models have been proposed, empirically derived from observations. {For example, \cite{Belov2015}, with their equation~2,} provides the variation of GCRs in a MC taking into account different contributions, including that of the magnetosphere.  {\cite{Masias2016}, with their equation~4, rather link the two-step FD to the variations of the} intensity and the fluctuations of the magnetic field within the substructures of ICMEs {containing MCs}.
{Another sub-structure of a typical FD is its frequently observed long recovery phase \citep[e.g.,][]{Usoskinetal2008}, but the precise main mechanisms that cause this are not yet well understood.}


Previous studies have either focused on the effect of the different substructures on the FD from case studies, or via statistical studies mixing different types of ICMEs (with or without a shock, with or without a MC, with different speeds). This has led to an ongoing confusion as to what are the basic ingredients that fundamentally drive the FD. We therefore propose to revisit the results of \cite{Masias2016} by further investigating the role played by ICMEs substructures, as follows: our dataset has been extended to include a revisited list of ICMEs spanning 17 years of ACE data (from 1998 to 2015), with associated neutron monitor data {to infer the GCR flux}.  While the previous study investigated the role of 44 ICMEs with MCs, we extend our list to more than 300 ICMEs with a clear ME, but not necessarily a MC (Section \ref{sec:datasets}). {These events are carefully selected from a larger list of events, in particular to study only isolated ICMEs.} We then investigate more precisely the difference between ICMEs with and without a sheath, and investigate the role of the magnetic field by selecting ICMEs with similar speeds to minimize the speed effect 
(Section \ref{sec:ejecta_without_sheath}). In Section \ref{sec:fd_nofd}, we further investigate the role of the magnetic field by comparing ICMEs that drive or not a FD. 
Finally, we conclude on the present study in Section \ref{sec:conclusion}.

\section{Datasets and method description} \label{sec:datasets}

\begin{figure*}[ht!]
\plotone{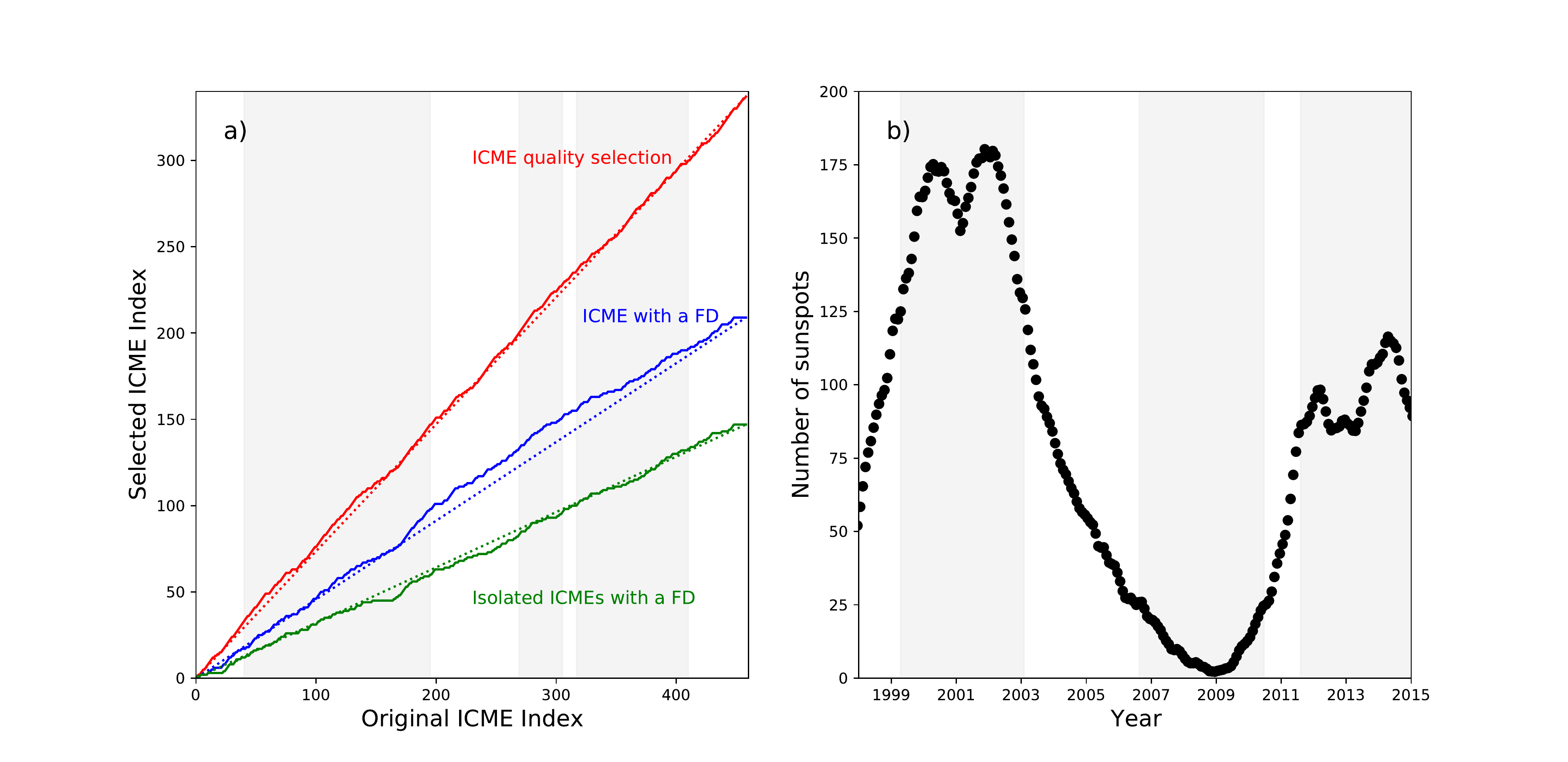}
\caption{Selection of events from the original list. Left: three selections of ICMEs are shown by their index number ($y$-axis) against the original index number ($x$-axis). The red plot (continuous line) shows all the ICMEs of good enough quality selected from the original dataset. The blue plot shows all the ICMEs with a consistent FD, and the green plot shows all the ICMEs with a FD that are isolated from surrounding FDs or GLEs. The dotted straight lines in the respective colors indicate the linear tendency (joining the first and last points of the data) for all selections, which provides a guide for the eye whenever there is a departure from the original list. Right: Monthly mean total sunspot number for the same year range as the ICME list. The shadowed areas correspond to the solar cycle maxima and minima, and are reported on the left graph with the equivalent ICME index numbers.
\label{fig:selection}}
\end{figure*}

\subsection{ACE data for the ICME list} \label{subsec:icme_dataset}

   
To investigate the properties of the plasma and magnetic field for ICMEs related with FDs, we use the Advanced Composition Explorer (ACE) space mission Level 2 mission data\footnote{\url{http://www.srl.caltech.edu/ACE/ASC/level2/index.html}}. ACE was launched on 25 August 1997 and has continuously provided solar weather monitoring from the L1 Lagrange point \citep{Stone1998}. In particular, we use the {magnetometer (MAG)} instrument that provides information for the interplanetary magnetic field \citep{Smith1998}, and the Solar Wind Electron, Proton and Alpha Monitor (SWEPAM) instrument for the proton speed, density and temperature \citep{McComas1998}. 

   
We also used the Richardson and Cane ICME list\footnote{
\url{http://www.srl.caltech.edu/ACE/ASC/DATA/level3/icmetable2.htm}} 
 that we revised, as reported in \citep{Regnault2020}. This revision includes checking/redefining the frontiers of ICMEs and only taking into account ICMEs with clear signatures. The full list of events can be found at \url{https://idoc.ias.u-psud.fr/sites/idoc/files/CME_catalog/html/}. 
For the purpose of our study, we revisited that table again, by removing all ICMEs that are in interaction with another ICME, or another solar wind structure (e.g. 
CIRs). Similarly, ICMEs with complex structures, or with boundaries difficult to estimate, are also not taken into account. We also removed from our sample those ICMEs for which more than 50\% of the data is lacking within each substructure (sheath and/or ME). From the original sample that contains 458 events, we found 337 remaining events that satisfied our quality criteria.


To check that there are no biases in the event flagging process (e.g. getting used to a certain type of structure which could then be more frequently selected, or a lassitude effect in the selection process due to the time it takes to scan the full database) we compare the ICME indexes as follows. 
At each step, the selected ICMEs are numbered by growing observed time order. This defines an ICME index (going from 1 to 458 initially, and from 1 to 337 in the final list). At each step, the scan of events is done with a growing index.
In Figure \ref{fig:selection}.a, we plotted the selected ICME index
(red continuous line) versus the original ICME index in our ICME sample table (on the $x$-axis). A red dotted line is over-plotted so as to show the linear trend. The idea is as follows: 
{for a fixed sample,}
if there is a bias that increases during the event scan (so by increasing ICME index, reported on the $x$-axis), then the slope of the graph would continuously {decrease/increase (depending on the bias sign).}

Similarly, if there are intervals of time when the flagging is more important than others, the slope of the graph would change within that interval of time. One can see that each increment of selected ICMEs follows the linear trend, meaning that there is no particular bias appearing. 


In Figure \ref{fig:selection}.b, we reported the number of sunspots from the SILSO international database\footnote{ \url{http://sidc.oma.be/silso/datafiles}} against the years covering our ICME list. We also indicated in gray shadows particular epochs in the solar cycle, namely the maximum of solar cycle 23 around year 2001, the minimum {in between cycles 23 and 24} around year 2009, and the maximum of solar cycle 24 around 2014. These areas are also over-plotted in Figure \ref{fig:selection}.a to check if there is any 
{dependency of selected cases} associated with these time periods. 
The first and third gray areas are broader, and the second gray area smaller in panel a) compared to b) since the cycle maximum/minimum contains a larger/smaller number of ICMEs, respectively. 
There is no significant departure from the linear trend {(red plot)} during these time periods, showing that no bias is introduced by a variable frequency of ICMEs.

\subsection{Forbush decrease event selection} \label{subsec:fd_dataset}


To investigate the FD events, we use two neutron monitors with equivalent rigidity cut-offs but situated at different positions on the globe. The McMurdo station is situated on Ross Island in Antarctica, while the Oulu neutron monitors are placed in the northern hemisphere in Finland. Due to the shielding of the Earth's magnetic field, McMurdo has a {geomagnetic} rigidity cut-off of $\sim$0.3 GV, while Oulu's is at $\sim$0.8 GV, while both monitors are at the comparable altitude from the sea level {(respectively at elevations of 48 m and 15 m\footnote{\url{https://www.nmdb.eu/nest/help.php}})}. The data have been retrieved from the the NMDB Real-Time Database for high-resolution Neutron Monitor measurements\footnote{\url{https://www.nmdb.eu}}.  
The choice of these monitors is also predicated by the availability that needs to start at least from the launch of the ACE mission to provide a complete dataset. This allows us to compare with the ICME list. The interval of data accumulation is hourly from 1960 (1964 for Oulu), and becomes a 1-minute interval data from 1998 for McMurdo (1996 for Oulu).

We only selected events corresponding to the refined list of ICMEs, for which there was no data gaps for the McMurdo neutron monitor, and that also had a visible FD. The selection for the existence of a visible FD is made by eye for each individual case. We therefore found a total of 209 events that were related with a good quality ICME and with a good FD event, while keeping in a separate table the 108 ICMEs that did not drive a FD. 
In Figure \ref{fig:selection}.a, the selection of these events is indicated with the solid blue line, and the linear trend with the dotted line. One can see that the selection is fairly linear, with 
a few more ICMEs selected as good events with respect to the typically expected average, as indicated by the blue dotted line, during the declining phase of the first shown solar maximum (around 2003).


In several cases, ICMEs are not isolated, so that a FD occurs before and/or after the ICME start/end because of the presence of compression regions related to CIRs or another ICME. 
Also, in some cases, the level of the GCR flux increases before the beginning of the FD associated with an ICME.
The necessity to remove these events comes from the fact that without this step, the superposed epoch analysis used in the following would be contaminated as these events do not express the direct answer of GCR flux produced by an ICME itself, but the answer is due to the combination of two consecutive events.
So as to not decrease the sample of ICMEs too much, we kept in the case of interacting ICMEs the ones with a weak preceding ICME that did not significantly affect the GCR level (less than 1\% of the background level). If an event after an ICME (i.e. a following ICME or CIR) affects the recovery phase of the event under study, for example by inducing another FD, the wake region of the ICME of interest is treated as a data gap. 


In Figure \ref{fig:selection}.a, similarly as above, we checked the removal of all these cases. This is shown with the green continuous line, and correspond to a sample of 147 remaining events. This selection therefore corresponds to a third of the original catalog. However, the linear trend shows that there is no particular period of time where the above selections of events is stronger, 
as this would have showed up as a non-linear trend in the graph. This shows that while the actual number of FD events vary during the solar cycle, their association with ICMEs already limits the sample number that we have: ICMEs occurring during the solar maximum are generally of lower quality, mainly because their frequent occurrence leads to interactions and therefore signatures that are more difficult to assess than isolated cases that are more numerous during the cycle minimum. 


As a final step, we cleaned all the data from Ground Level Enhancements (GLEs), which are induced by large solar particle events with particles energetic enough to propagate through Earth's magnetosphere and atmosphere and eventually trigger an enhancement in such ground-based detectors.
First, we automatically set an automatic threshold for the neutron monitor values.  This threshold{, of value 120 \%,} is fixed at a level that is higher than the usual fluctuations of the GCR flux. Every time the GCR flux level crosses this threshold, the values are automatically assigned to a data gap.  As a next step, we also control all the results by scanning visually the remaining NM data and we remove the remaining GLE intervals for which the GLEs were too weak to be detected automatically.
The time intervals corresponding to GLEs are much shorter in duration than FDs, therefore leading to a few data gaps. Therefore, this step does not affect the count of the total number of FD events. 


We therefore end up with the following two samples that fulfill the quality selection: a sample of ICMEs with a FD that is isolated from any other event, and a sample of ICMEs with no FD.
Confident in the selection of our FD-inducing ICMEs, we investigate in the following the generic properties of the FD profiles on the sample of carefully selected ICMEs and FD events.

\subsection{The superposed epoch era method applied to GCRs} \label{subsec:method}

   
The superposed epoch analysis \citep[or Chree analysis,][]{Chree1913} consists in normalizing a dataset, commonly a time series of different events on a similar timescale, and obtaining the time series' behavior of a statistically relevant parameter for each normalized time bins. In the context of ICMEs, {previous studies \citep{Lepping2003b, Rodriguez2016, Masias2016, Janvier2019, Regnault2020}} typically take well-defined times for the normalization: the time of the discontinuity, the start time of the ME or MC, and the end time of the ME or MC. In the following, the normalization is similar to the studies of \cite{Masias2016, Regnault2020}: the unit time for the normalized time series is taken as the length of the sheath (if it exists), and the ME is normalized to three times this unit time. {This choice is motivated by the previously cited studies, which observed that sheaths have on average a duration three times less that of the ME. However, we point to the reader that the choice of the substructures lengths for the SEA is mainly done for visualization of the results, since each are independently constructed via the normalization and binning steps. Then, i}n the case of ICMEs with no sheath, {the time binning of the ME is the same as ICMEs with a sheath, so that} the ME is still normalized to three times the unit time. This renders a direct and simpler visual comparison of the graphs.
Then, all the events are binned with an equal number of bins within the sheath, the ME, the solar wind before the ICME and the solar wind after the ICME (i.e the wake).


As we are interested in looking at the variations of the GCRs long before and after the ICME, we decided to take for both the pre-/post-ICME solar wind length 3 times the time duration of the ME. This is especially important to catch enough time for the recovery phase, i.e. the time it takes for the GCR level to come back to the pre-ICME levels. Of course, since the GCR intensity level fluctuates within the solar wind (with an overall fluctuation following the solar cycle), the superposed epoch analysis is only consistent if the background intensity level is normalized. Therefore, for each event, we normalize the GCR intensity with the averaged level in the pre-ICME solar wind, (i.e. 3 times the length of the ME for an interval considered before the shock/discontinuity for cases with a sheath, or before the ME start for those without).
The GCR flux in each superposed epoch is then given as a percentage of this averaged background level. 

\section{Can magnetic ejectas without sheaths drive a FD?} \label{sec:ejecta_without_sheath}

In the study led by \cite{Belov2015}, the authors showed that MCs can drive FDs. This was also discussed in the simulation paper by \cite{Benella2020} where the authors investigated the influence of the large-scale magnetic field in driving FDs. However, in the sample from \cite{Belov2015}, the MCs were clearly associated with a sheath, driving a FD before that of the MC (e.g. their Figure 5).
Furthermore, we showed in \cite{Regnault2020} that the magnetic field and plasma profiles of the ME were somewhat different between ICMEs with a sheath and those which did not drive a sheath.  ICMEs not related with a sheath are generally slow ICMEs which were not able to pile up enough solar wind material at their front, while still retaining a coherent magnetic structure allowing their detection as a clear ME.  As such, we investigate in the following the impact of the existence or the absence of a sheath on the possibility to drive a FD, and its characteristics, via the SEA.
We found in our sample 130 cases of ICMEs that have a well defined sheath, and only 17 cases of ICMEs without a sheath that are able to drive a FD {(for a total of 147 cases)}.

\begin{figure*}[ht!]      
\plotone{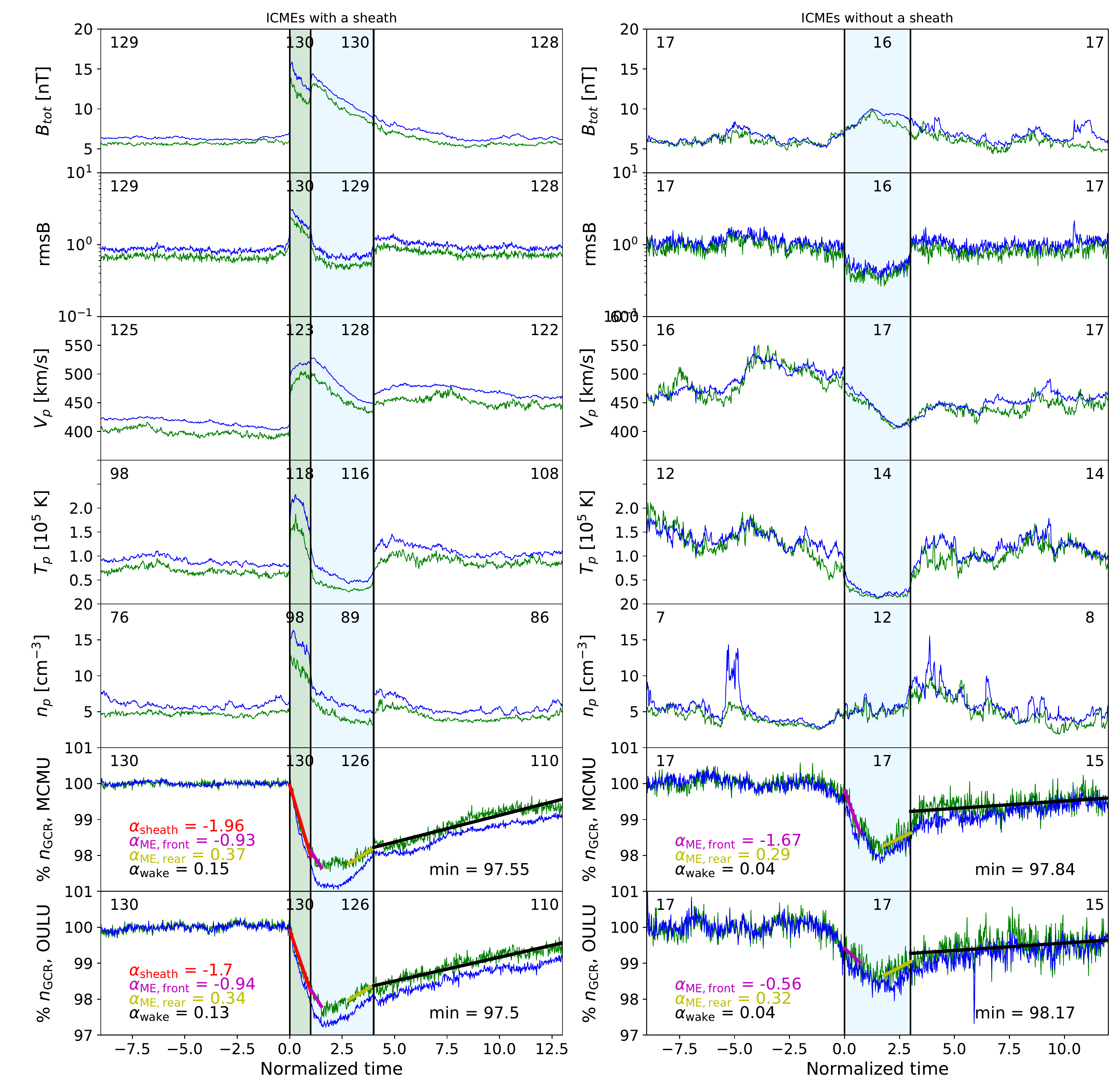}
\caption{Superposed epoch analysis for the 130 ICMEs with a sheath (left) and the 17 ICMEs without a sheath (right), in our selection of events that are associated with a FD. We plot the total magnetic field intensity (1st row), the magnetic field fluctuation magnitude $rmsB$ (2nd row), the proton speed \Vp\ (3rd row), the proton temperature \Tp\ (4th row), the proton density \np\ (5th row), and the GCR flux in percentage (normalized to the pre-ICME solar wind) for the McMurdo and Oulu neutron monitors (6th and 7th rows). The median is in green line, and the mean in blue line. The sheath substructure is indicated with the shaded green area, while the ME is indicated with the blue area. {The number of events is written at the top for each substructure region.}
Finally, for the GCR data, we also add the linear fit of the median to the different FD phases (decrease in the sheath with the red linear fit, in the ME front in magenta, the recovery in the ME rear in blue, and the recovery in the wake in black). The associated slopes for the linear fit are also indicated, with the same color code. The minimum values indicate the minimum reached by the respective colored curves. 
}
\label{fig:icme_ws_ns_fd}
\end{figure*}

\subsection{ICMEs with or without a sheath driving a Forbush Decrease}
\label{subsec:icme_ws_ns_fd}


The results of the SEA for both categories are given in Figure \ref{fig:icme_ws_ns_fd}. 
For each substructure (pre-ICME solar wind, sheath, ME and wake), we indicate the number of events that were taken to build the superposed epoch. For each substructure, we only take the events that display enough data (more than 30\% of the substructure length). 


Comparing both SEA for ICMEs with and without a sheath first shows that because of the scarcity of the events for ICMEs without a sheath, all parameter plots are much noisier than the SEA for ICMEs with a sheath. This still allows us to define a median (in green) and a mean (in blue) and they are closeby since the ICMEs without sheath have not the long tail of fast events (present for ICMEs with a sheath).  Next, for both ICME groups, the magnetic and plasma profiles follow the same trends as already described in details in \cite{Regnault2020} (see their Figures 2, 3 and related text). 
We recall the most important information here, which is that: 

\begin{itemize}
    \item The magnetic profile of the ejecta is completely asymmetric for the ICMEs with a sheath, with a decreasing slope starting from the front to the rear of the structure, while it is almost symmetric in the case of the ME without a sheath. Indeed, when a sheath is present, its extra pressure compress the ME front and this transforms the B strength profile in the ME, as was discussed in the \cite{Regnault2020} study.
    \item Both the absolute and relative speeds of the ME are higher for ICMEs that drive a sheath (where the relative speed means the difference between the ME mean speed and that of the solar wind in front of the ICME).
    \item {The magnetic field strength, its $rmsB$ (which measures the fluctuations of the magnetic field), the plasma temperature and density are strongly enhanced in the sheath compared to the pre-SW values}. 
\end{itemize}


With this in mind, we then focus the discussion on the GCR flux (second to last and last rows). The evolution of the GCR levels follow the same trend for both McMurdo and Oulu neutron monitors{, with a lower maximum decrease for McMurdo, as expected for the relative rigidity cut-off of the two locations}.
For both, we find that while ICMEs with and without a sheath drive a FD, the decrease is slightly larger for ICMEs with a sheath. Indeed, comparing the median curves (in green) and the mean curves (in blue), we find that the minimum of the decrease reached for ICMEs with a sheath is approximately 3\% of the background value in the McMurdo data for the mean and 2.5\% for the median (2.8\% and 2.5\% in the Oulu data, so similar values).  
For ICMEs without a sheath, the minimum reached for the mean curve in the McMurdo data is 2.3\% and for the median 2.2\% (2\% and 1.8\% in Oulu data{, so similar values). We expect similar behaviors for the Forbush decrease observed at both McMurdo and Oulu monitors, due to the fact that they both cover the same range of GCR energies and generally show similar amplitudes for the FD \citep[see e.g.][]{Thomas2014}.}
In the case of ICMEs with a sheath, the mean and median curves show a larger difference, which is expected since the mean is more affected by large events (which all have a sheath).  
We conclude that comparable result for the two neutron monitors gives us the confidence that the results seen are not just related to random fluctuations of one neutron monitor and/or choice of site, as both are situated in the two different hemispheres. 


The slopes indicated in the figures are calculated in the normalized time frame of the superposed epoch eras. 
We indicate them in the graph, however because the normalization is made on the basis of an average duration of the sheath and the ME, we do not compare directly the slopes between the different substructures. The idea here is that the SEAs are representative of a generic ICME, and indicating the slope here gives an indication of the tendency of the decrease in different substructures. However, to compare quantitatively the dispersion of the slopes of the FDs, one would need to go back to the whole sample and investigate these slopes for each event, which is beyond the scope of the present paper.


Looking at the recovery part of the FD, we observe several points. First, the recovery phase within the ME (rear part, yellow slope) is similar whether or not there is a sheath. The recovery is however much more gradual in the solar wind in the wake of the ICME for those that do not have a sheath (slope of 0.04 compared with $\approx 0.14$). Second, for both cases, even after 3 times the ME duration, the solar wind has not recovered the original GCR level (by about 1\%). 


These first results point that ME can also drive a strong FD even if it has no sheath at its front.
\cite{Kumar2014}, in their study, found that ICMEs that have a shock are more likely to drive an intense FD that ICMEs that do not (although we argue here that more than the presence of a shock, it is the sheath that drives the decrease in the GCR count). We find here that the decrease is actually of comparable intensity (within 0.5\% difference in the minimum reached for the FD for both categories). Furthermore, the profile of the FD in the ME with and without a sheath is reminiscent of the results of \cite{Belov2015}: slow MCs tend to drive FDs which minimum is at the center of the symmetric structure, while fast MCs (and therefore typically associated with a sheath) show an asymmetric FD which minimum is closer to the front edge of the MC.

\begin{figure*}[ht!]          
\plotone{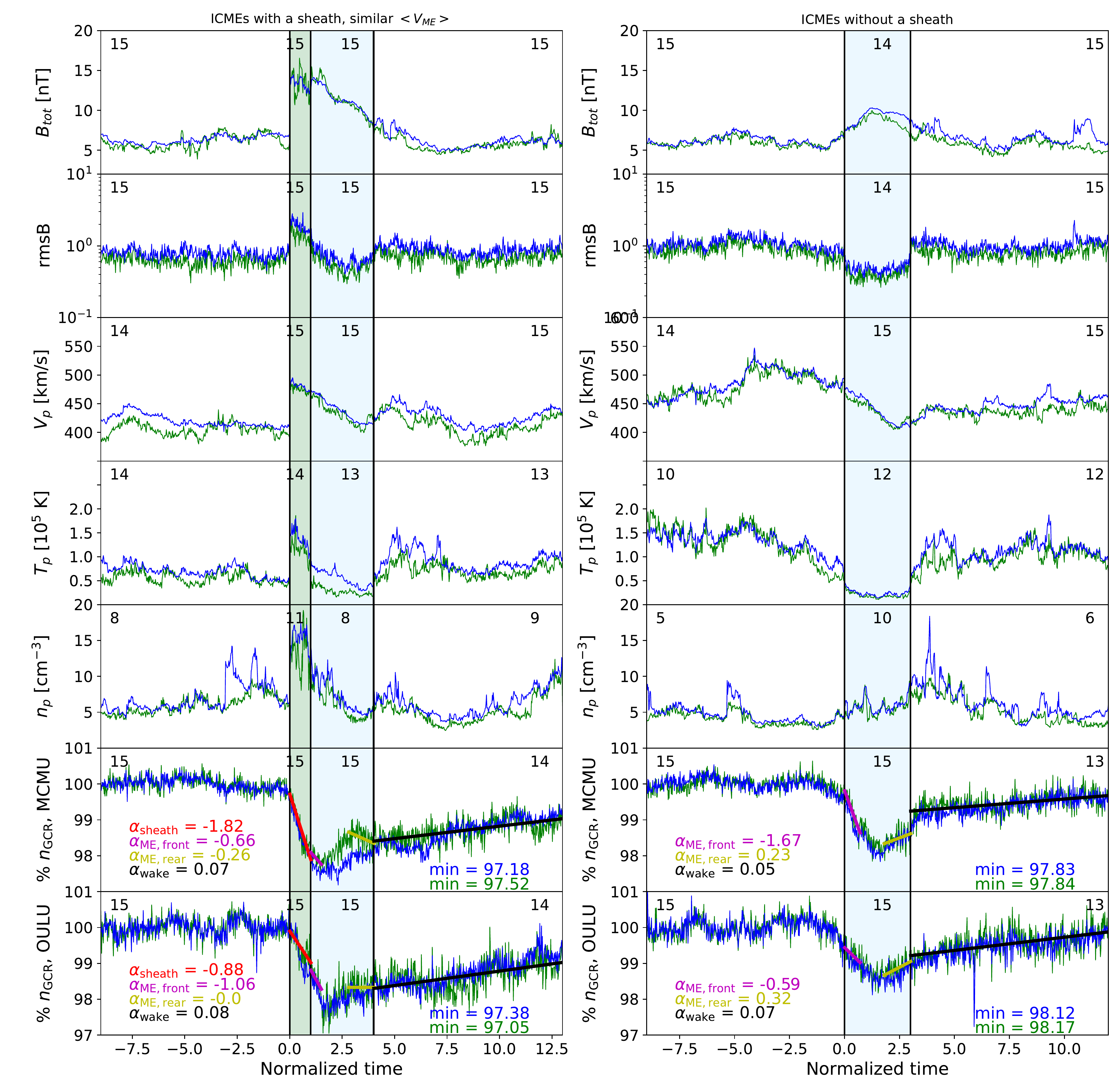}
\caption{Superposed epoch analysis for the 15 ICMEs without a sheath (right) and the closest ME mean speed ICMEs with a sheath  (left). 
The parameters and the drawing convention are the same as in Figure \ref{fig:icme_ws_ns_fd}.}
\label{fig:icme_ws_ns_fd_speedcomparison}
\end{figure*}

\subsection{Same speed effect for ICMEs with and without sheaths}
\label{subsec:icme_ws_ns_fd_speedcomparison}

In \cite{Masias2016}, the authors showed that a superposed epoch analysis made on a whole set of ICMEs, mixing different aspects (i.e. different averaged magnetic field intensities, speeds, ...) hides some features that are associated with intrinsic properties such as the absolute speed of the event (calculated from the mean speed of the ME). Hence, an SEA made on categories of ICMEs (i.e. ranked by speed) better highlights the relation between different properties.
To minimize the effect of both the statistical weight and the speed parameter when comparing ICMEs with and without sheaths, we search the pairs in our ICME lists with and without a sheath that have the closest average speeds $<V_{ME}>$, within 10 km/s of difference.  We found 15 pairs for 17 ICMEs without sheath (we remove 2 associations that were duplicating the same associated ICME with a sheath). 
While within the two samples, we still have ICMEs with different parameter profiles, both samples now display ICMEs selected with similar speeds so that the effect is overall the same for both samples, allowing comparisons between them.  The results of the SEA on these events is reported in Figure \ref{fig:icme_ws_ns_fd_speedcomparison}.


Because of the same number of events, we have now the same statistical noise for all parameters (up to fluctuations in the ICME numbers due to data gaps). 
In these graphs, we see that {both proton temperatures and densities, inside the ME, have similar values for MEs with/without sheaths.}
We remark that the ME magnetic field profile of the ICMEs with a sheath is less asymmetric, with a less pronounced sheath than when the whole sample is taken (Figure \ref{fig:icme_ws_ns_fd}).  This is understandable, since we have selected ICMEs with slower speeds, which are linked with weaker sheaths, so with smaller effects on the following ME \citep[see results from][]{Masias2016, Regnault2020}. 

Next, while we selected the ICMEs with a sheath sample only with a criterion on the mean speed, we ended up with a similar speed profile in the SEA. Therefore, the cases with and without sheaths are even more comparable with the same self-similar expansion profile.  This result is a consequence of the ICME expansion rate being driven by the decrease of the surrounding solar wind (hereafter, SW) pressure with solar distance \citep{Demoulin2009,Gulisano2010}.  This decrease is dominated by the density decrease, which is constrained by the mass conservation, so that the SW pressure scale with a similar power of distance in various SW.  It results in an ICME difference of velocity between the front and the rear which is mostly defined by its mean velocity, then the selection of this velocity also implies a similar velocity profile. 

For the ICMEs with a sheath, this linear velocity profile extends in the sheath {\citep[a property of slow events, see][]{Masias2016}}.  Then, as expected, the case of ICMEs with a sheath have a higher speed compared with the ambient pre-ICME solar wind (which is at the origin of the sheath formation).  This is in stark contrast with ICMEs without a sheath, where the speed in the pre-ICME solar wind is much higher, so that no sheath could form. Therefore, ICMEs without a sheath, while having a clear ejecta (as seen in the magnetic field, rmsB, speed and temperature profiles), move in a faster solar wind that peaks at around 500 km/s. 


Analyzing the FD in both McMurdo and Oulu data (the two last rows of Figure \ref{fig:icme_ws_ns_fd_speedcomparison}), we notice that the slope in the sheath for the selected ICMEs with a sheath is lower ($s_\mathrm{MCMU}=-1.82${($\pm 0.12$)},  $s_\mathrm{OULU}=-0.88${($\pm 0.13$)}) than for the full set ($s_\mathrm{MCMU}=-1.96${($\pm 0.07$)}, $s_\mathrm{OULU}=-1.7${($\pm 0.07$)}). {We note that the values of the slopes are significantly different for the Oulu monitor, while appreciating that the reduced sample of MEs with a sheath is quite small and may explain the comparable values found {for McMurdo monitor}.} As above for the magnetic field and plasma parameters, this is a consequence of selecting on average slower ICMEs (with sheaths).
However, the maximum decrease attained by the FD is of similar amplitude as in the whole ICMEs with a sheath sample. This points to the fact that the decrease continues in the ME.


As found previously \citep[][ and references therein]{Masias2016}, we confirm that the sheath is associated with a higher rmsB level 
as shown in the second row of Figure~\ref{fig:icme_ws_ns_fd_speedcomparison}. In contrast, the ME, both with and without a sheath, has a clear decrease in rmsB.  However, both the sheaths and MEs without a sheath drive FDs of similar intensities, as can be seen in both McMurdo and Oulu neutron data. Then a FD can be observed without an enhanced $rmsB$.  This indicates that the level of magnetic fluctuations has a small effect in screening GCRs. {The same conclusion is reached for the plasma parameters ($T_p$ and $n_e$}) which, as $rmsB$, are only enhanced in the sheath.


The results of the bottom panels of Figure \ref{fig:icme_ws_ns_fd_speedcomparison} show that in the presence or an absence of a sheath, the ME is still driving a diminution in the count of GCRs. {The amplitude of the Forbush decrease from the beginning of the ME to the minimum level reached is smaller in the McMurdo data (0.7\%) when there is a sheath at the front of the ME compared with ME with no sheath (1.82\%), although these amplitudes are comparable in the Oulu data.} {Because of the low number of events in the sample with no sheaths, it remains difficult to compare the amplitude of the decrease within the ME for both categories. 
} Since MEs that are preceded by a sheath have a higher magnetic field, so a priori, a potentially more effective screening mechanism, this points to the fact that the sheath already provides a screening of GCRs before the ME starts to play role.  This result explains why the FD is less marked in the ME for ICMEs with a sheath. However, the fact that the FD occurs within the ME also indicates that the sheath is not 
the only structure responsible for the GCR decrease. 

Since the magnetic modulation of charged particles is energy-dependent (with low-energy particles more easily affected), the ``screening process'' preferentially starts with GCRs with lower-energies. With the presence of a sheath and its deflection of GCRs, the remaining GCR flux has a larger proportion of high-energy particles which are more difficult to be modulated by the magnetic fields of the following ME.
We conclude that the sheath rather screened most of the GCRs that could be screened, so those of lower energy, while the remnant bulk of GCRs would require a higher magnetic field to be screened.  In contrast, when no sheath is present, the screening is done by the ME with a slightly weaker magnetic field compared to the sheath (about 8.6 nT on average compared to 11.1 nT). 


The present results clarify those found in \cite{Belov2015}, which stated that FDs driven by fast MCs are asymmetrical, with a higher FD magnitude and a minimum that is closer to the front edge of the MC. What we find here is  that all MEs, and in particular MCs, have the potential to drive FDs.  But for fast ICMEs, the FD is overshadowed by the presence of the sheath that drives most of the GCR intensity decrease, so that the FD minimum is closer to the ME front. Finally, the contribution of the magnetic field in fast MEs is weak, and its asymmetry is not revealed in the FD profile at the level it would be without a sheath.

\subsection{Recovery in the ICME wake}
\label{subsec:recovery-wake}


We also calculate the recovery slope within the wake (i.e. the solar wind post-ICME, Figure \ref{fig:icme_ws_ns_fd_speedcomparison}). We find that the slopes are similar (0.05 to 0.08), so that the presence or not of a sheath has only a weak effect on the recovery phase. 
This result is in sharp contrast with the previous result of Figure \ref{fig:icme_ws_ns_fd} where the slopes were 0.14 (mean of McMurdo and Oulu) and 0.04, respectively.  This again emphasizes the importance of the ICME velocity which is larger in the results of Figure \ref{fig:icme_ws_ns_fd} compare to Figure \ref{fig:icme_ws_ns_fd_speedcomparison}.
Finally, three times the interval of the ME length {allows to recover only about 3/4 of the FD minimum towards} the original level found in the pre-ICME. The long-lasting effect of the ICME passage on the solar wind was discussed in previous papers \citep{Temmer2017, Janvier2019}. Here, we find that this effect is also seen in the FD recovery.

  
We next analyze this long recovery phase. 
The sheath and/or the ME act as a screening body that removes a certain amount of the GCR fluxes coming from the outer space (at distances above 1 au). After the passage of the ME, the solar wind in the wake is still under the shadow effect of the sheath and/or the ME at its front {(while no jump in the GCR flux is present at the shock, so its screening is negligible )}. 
This solar wind is therefore associated with the contributions of the GCRs that arrive from behind the ME. 
This shadow effect decreases as the solid angle 
made by the sheath/ME decreases away in the ME rear. 
Then, the recovery involves the evolution of the global screening of the ICME with distance to the ICME rear.  


This screening is also a priori expected to be at work in front of the ICME, with a removal of part of the GCR flux coming from the Sun side. Such a decrease of flux is observed only in front of a small fraction of ICMEs, but it is not a general feature of ICMEs with sheath, while a weak GCR flux decrease is present before ICMEs without sheath (left and right columns of Figure \ref{fig:icme_ws_ns_fd}, respectively).
{Indeed, for ICMEs without a sheath, there is a slight decrease of about 0.3 \% at most for the GCR flux in the pre-solar wind. This decrease} starts at about $t_{\mathrm{norm}} \approx -2$, around the same time as a slight increase of $B$.  This location also corresponds to a faster pre-SW which shows an expansion as the profile of $V$ decreases with time. This can contribute to a slightly lower GCR flux by dilution, since the rate of GCRs diffusing into the magnetic structure is not high enough to compensate the density decrease due to the ME expansion.
Then, a local, rather than global, screening effect is expected to create this GCR flux decrease before ICMEs without sheaths.

\begin{figure*}[ht!]          
\epsscale{1.2}
\plotone{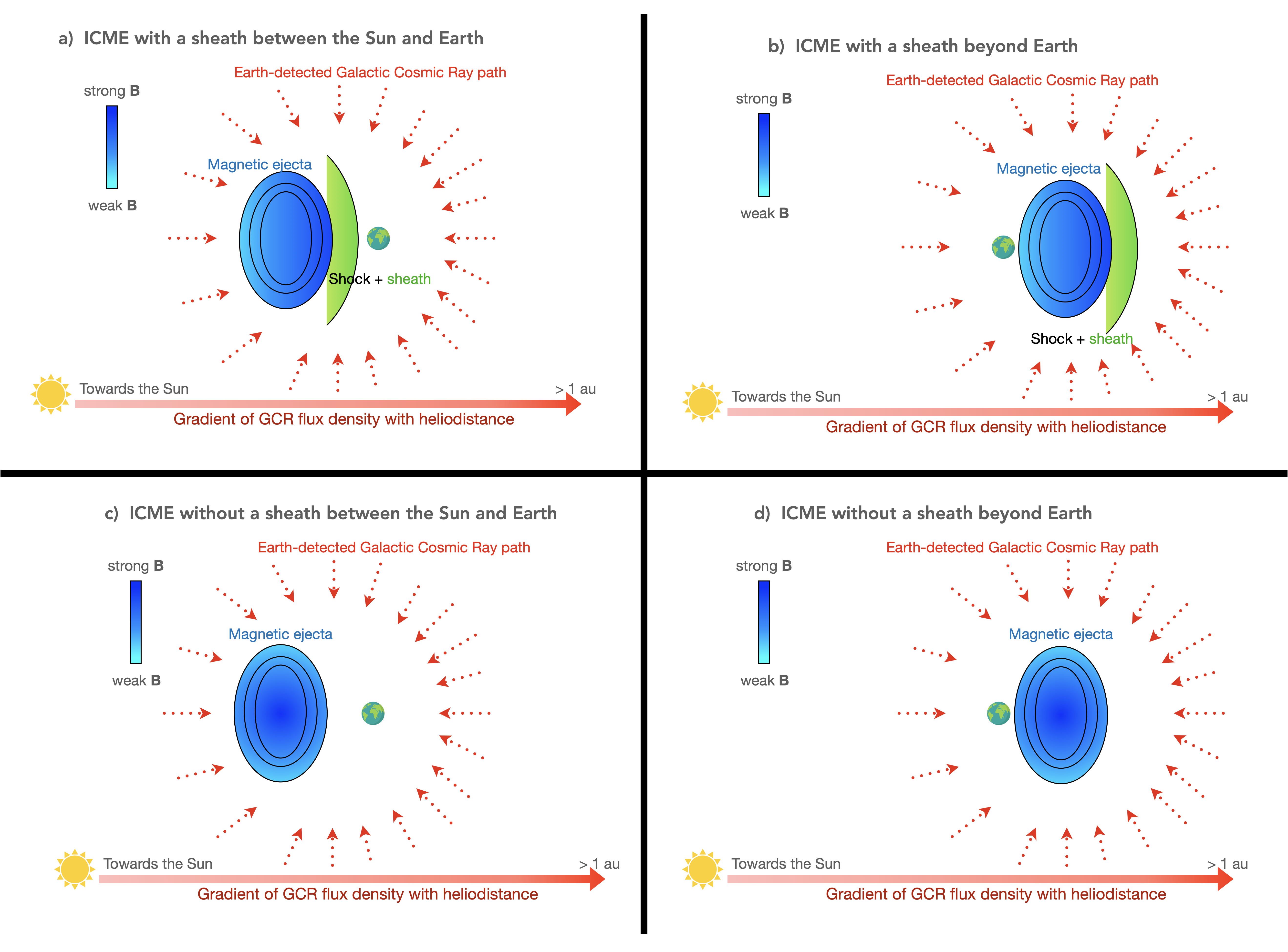}
\caption{Cartoon summarizing three contributions to the anisotropy of the GCR seen before and in the recovery part of the Forbush Decrease. Panels a) and b) represent the passage of an ICME with a sheath, with a magnetic ejecta with a stronger magnetic field at the front (closer to the sheath region, in green) than at the back. Panels c) and d) represent the passage of an ICME without a sheath, with a generic magnetic ejecta with a symmetrical magnetic field profile. The GCR flux density is represented by the arrow density.  The anisotropy observed in the SEA is most likely due to the gradient of the GCR flux with heliodistance, since the anisotropy is seen in ICMEs without a sheath (c and d).
}
\label{fig:cartoon-gcranisotropy}
\end{figure*}

 
We next analyze the plausible origin of the front/rear asymmetry of GCR flux in ICMEs.
A first source of  screening asymmetry is the sheath location. Indeed, the GCR profile is more symmetric within ICMEs without sheath (right column of Figure \ref{fig:icme_ws_ns_fd}). However, the presence or not of a sheath has only a weak effect on the recovery phase when ICMEs of similar speeds are compared (Figure \ref{fig:icme_ws_ns_fd_speedcomparison}).  Then, the contribution of the sheath to the asymmetry is not dominant in the wake.

A second asymmetry source is the magnetic profile, since the screening can be more efficient with a larger $B$ strength.  For ICMEs without a sheath, this asymmetry source is weak because the magnetic field profile is nearly symmetric in the ME.  However, we find that there is a front/rear asymmetry of GCR outside the ME, therefore not explained just by the profile of the magnetic field.

A third asymmetry source is the expansion of ICMEs with solar distance so that the screening solid angle and the radial extension of the ME are larger when the GCR flux is observed at the ICME rear than at the front region.  This effect is still weak however, considering the velocity profile observed for the ICMEs {shown in Figure \ref{fig:icme_ws_ns_fd_speedcomparison}. Indeed, we find that the velocity changes by less than 20\% across the ME.
In particular, we do not observe the consequence of such an  expansion on the magnetic profile for ICMEs without a sheath, which implies the same on its extension with the magnetic flux conservation \citep[see][for an analysis of expansion]{Demoulin2020}.}
Moreover, the evolution of these geometrical parameters are counteracted by a {weaker magnetic field (due to magnetic flux conservation), so a less effective screening with time.}  

Finally, the SEA of ICMEs without sheaths points to a dominant fourth source of asymmetry which is the intrinsic spatial asymmetry of GCRs density with a larger (lower) flux coming from the far (close) side of the Sun.  With this asymmetry dominating other processes, even the same ME screening can produce a much stronger GCR decrease at the rear than at the front of the ME.     We conclude that the spatial asymmetry of the GCR {density} is expected to be a key point in the presence of a long recovery phase at the rear of ICMEs compared to short and infrequently observed decrease of GCR flux in the front. Indeed, it has been shown that the intensity of GCRs is highly dependent on the radial distance from the Sun.  This was demonstrated by comparing Voyager and IMP measurements, which showed that the {GCR intensity is higher in the outer} rather than in the inner heliosphere \cite[e.g.][]{Heber2006}.  {This is a consequence of an increasing GCR shielding toward the Sun, in particular by the interplanetary magnetic field.} 

The contributions of all these, {except the expansion effect}, i.e. the presence of a sheath, the asymmetric magnetic field profile in the case of ICMEs with a sheath, and the intrinsic heliospheric radial asymmetry of GCRs {intensity}, are  reported in Figure \ref{fig:cartoon-gcranisotropy}. Because of the robust results found for ME without a preceding sheath, and their symmetric shape, the radial {dependence} of GCRs {density} seems to be the most likely candidate to explain the long recovery phase found in the present study. Furthermore, the level of asymmetry found here is in quantitative agreement with other studies that point to a GCR flux increase observed in different spacecraft measurements. For example, a $<10\%$ per au decrease was discussed in \cite{Lawrence2016}, measuring the gradients with MESSENGER and the Cosmic Ray Telescope for Environmental Radiation (CRaTER) on the Lunar Reconnaissance Orbiter, but also the 2\% to 4\% per au increase found in other studies beyond 1 au, e.g. \cite{Heber2002, Honig2019, Roussos2020}).
A quantitative modeling of this effect is worth to do, while outside the scope of present paper.


We conclude that {when ICMEs with and without a sheath produce FDs, both can drive FDs with} similar strength provided that the comparison is done for similar mean ME velocity {(which implies similar velocity profiles)}. The existence of a sheath implies only that the FD starts earlier with a large decrease than in the ME, but both ME preceded or not by a sheath can drive FDs with nearly equivalent minimum depletion.   When a sheath is present, most of the screening is made by the sheath, although the ME can also screen a bit further the GCRs. The recovery starts for both cases already within the ME.
With comparable ME mean speed, the recovery rate is comparable with and without a sheath, and about 3/4 of the recovery from the FD minimum is achieved after 3 times the ME duration.
Finally, we propose that the asymmetry of GCR flux at the front/rear of ICMEs without sheaths is dominated by the GCR angular anisotropy.

\section{On the importance of the magnetic field intensity} 
\label{sec:fd_nofd}


Following the previous results, we have shown that for equivalent speed profiles, MEs with and without sheaths can both drive FDs with comparable magnitudes. Since both sheaths and MEs have different magnetic fluctuation profiles, this parameter is not a necessary condition for the FD. 
In the following, to further investigate the role of the magnetic field intensity versus that of its fluctuations, we select equivalent ICMEs (with or without a sheath) that drive or not a FD. Those that do not drive a FD come from the original list (see Section \ref{sec:datasets}) for which no associated FD was found. Finally, we select pairs of events with equivalent speeds and magnetic field intensities, so as to remove the effect of these parameters. 

\subsection{ICMEs without a sheath}
\label{subsec:icme_ns_fdnofd_speedcomparison}

\begin{figure*}[ht!]             
\plotone{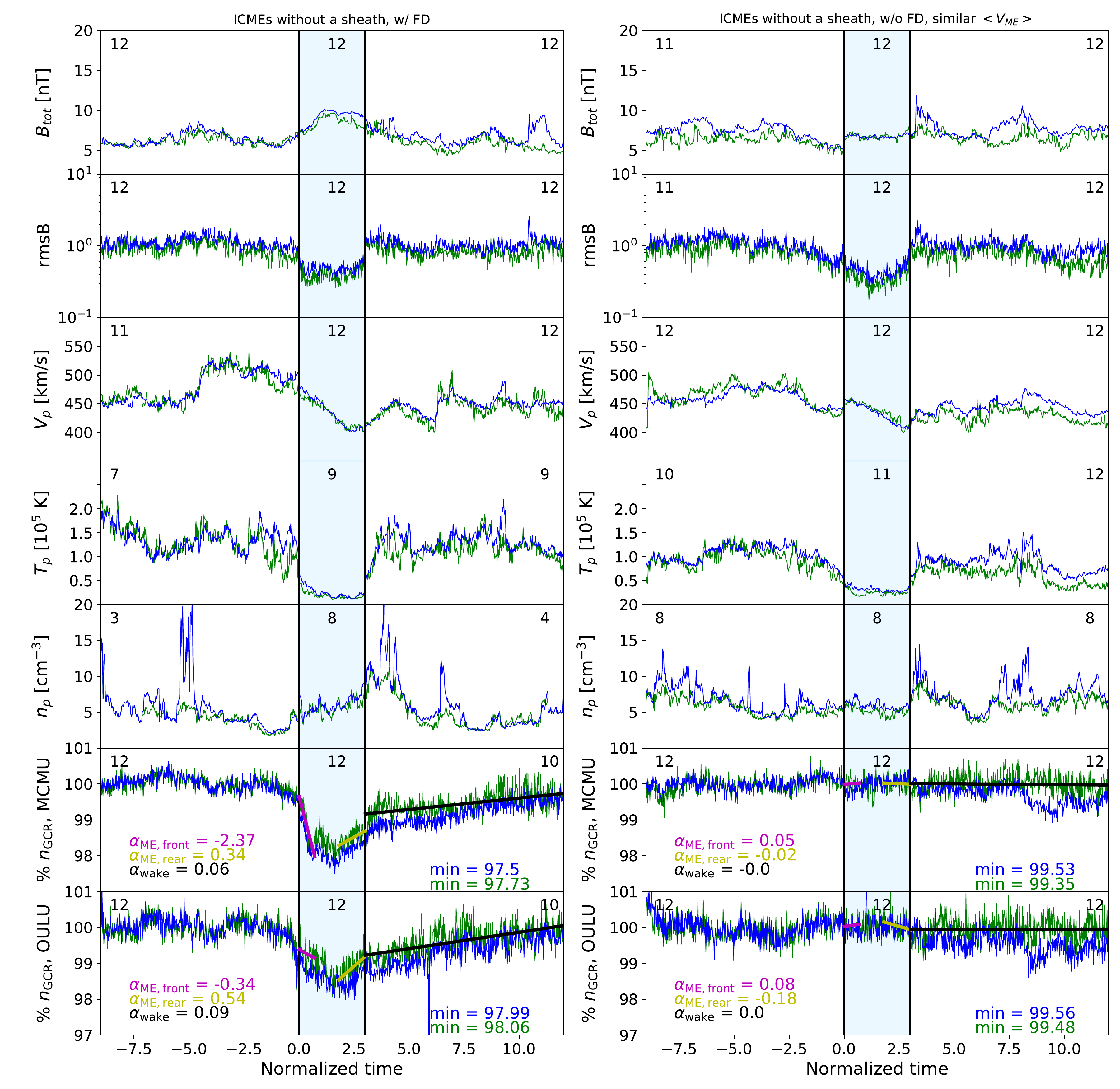}
\caption{Superposed epoch analysis for ICMEs without a sheath. Left column:  SEA on a sample of 12 ICMEs with a FD. Right column: SEA on a sample with 12 other ICMEs without a FD and that have the closest ME mean speed from the ones on the left side. 
The parameters and the drawing conventions are the same as in Figure \ref{fig:icme_ws_ns_fd}.}
\label{fig:icme_ns_fdnofd_speed}
\end{figure*}


In Figure \ref{fig:icme_ns_fdnofd_speed}, we first investigate the events categorized as ``ME only'' by selecting all the ICMEs with no detected sheath.  Similarly to Figure \ref{fig:icme_ws_ns_fd_speedcomparison}, we retain the couples of ICMEs with nearby mean ME velocity.
Looking at all the plasma and magnetic field parameters, we found no substantial differences within the ME in all the plasma parameters as well as in the the rmsB profile. The only main difference is that the intensity of the magnetic field is slightly higher in the MEs associated with a FD ($<B_{SW}> = 6.$~nT in the solar wind before the ICME compared with $<B_{ME}> = 8.5$~nT in the ME) while the MEs without FDs have a field strength comparable to the SW background ($<B_{SW}> = 6.3$~nT in the ME compared with $<B_{ME}> = 6.7$~nT). 


The present results confirm that the time variation of the GCR intensity during the GCR decrease is well correlated with that of the magnetic field \citep{Badruddin2016}. Since the level of the rmsB is similar in both cases, Figure \ref{fig:icme_ns_fdnofd_speed} also shows that the intensity of the magnetic field is the main driver of the FDs, provided the comparison is made on MEs of similar mean velocities. Therefore, the fluctuations of the magnetic field is clearly not the driver of the FD here, contrary to what was proposed by other authors \citep[e.g.][]{Arunbabu2015}.


Finally, for completeness, we compare the sample of ICMEs without a sheath driving a FD, with those that do not drive a FD, by selecting the closest mean magnetic field within the ME. Because the magnetic field profile and speed profiles are both different in the two samples, the results are less conclusive than the above selection. Since it is not possible to separate both contributions (speed and magnetic field), we leave this comparison to the Appendix \ref{sec:appendix} for reference.

\begin{figure*}[t!]           
\plotone{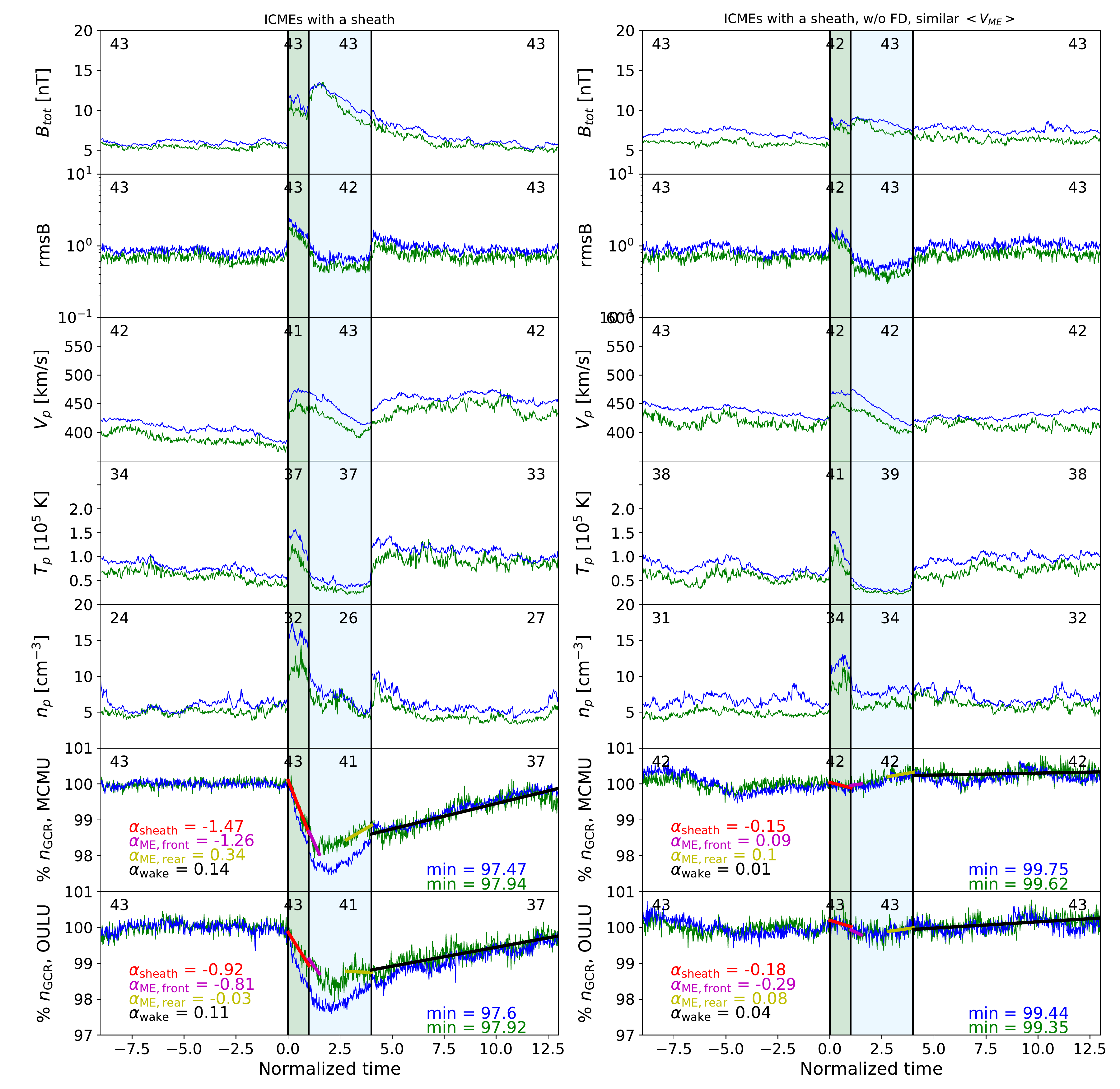}
\caption{Superposed epoch analysis for ICMEs with a sheath. Left: 43 ICMEs with a FD. Right: 43 other ICMEs without a FD and that have the closest ME mean speed from the ones on the left side.
The parameters {and the drawing conventions} are the same as in Figure \ref{fig:icme_ws_ns_fd}.}
\label{fig:icme_ws_fdnofd_speed}
\end{figure*}

\subsection{ICMEs with a sheath}
\label{subsec:icme_ws_fdnofd_speedcomparison}

We complete the previous subsection analysis by performing the same analysis with ICMEs having a well defined sheath. The selection is as before: to remove the effect of the speed, we only select pairs of ICMEs that have almost the same mean speed, narrowing our selection to 43 ICMEs in both lists, and the results are presented in Figure \ref{fig:icme_ws_fdnofd_speed}.


Here again, comparing the magnetic field and plasma parameters shows {that the magnetic field strength, plasma density and temperature are enhanced in the sheath of ICMEs with a FD.  However the high energy particles of GCRs have only a minor interaction with the plasma.  Apart from that, all plasma parameters, and rmsB have similar values in the sheath and ME for the two columns of Figure~\ref{fig:icme_ws_fdnofd_speed}}.  Then, we highlight again the main difference between the two samples as being in the intensity of the magnetic field.  For completeness, we show the same analysis by selecting the closest $<B_{ME}>$ in the appendix (Figure \ref{fig:icme_ws_fdnofd_mag}). We find that comparing events with close speed profiles provides the best comparison in a FD analysis.


We conclude once again that the magnetic field intensity seems to be the main driver of FDs, rather than the sheath turbulence and especially not the shock, as was often put forward in previous analyses. This result is obtained for ICMEs, where the only difference found when comparing ICMEs with a sheath and with similar speeds, that drive or not a FD.
We reinforce that the speed $V_p$ plays a key role in several ways: it indicates the overall travel time from the Sun, which would itself {(combined with the expansion rate)} be related to the time it takes for GCRs to diffuse into the magnetic structure (e.g. flux tube), as well as the time it can take for slower/faster structures to build a stronger sheath \citep[see for example the study of][]{Regnault2020}. By removing the dependency on the ME speed (as was done in Figures \ref{fig:icme_ns_fdnofd_speed} and \ref{fig:icme_ws_fdnofd_speed}), we are then able to pinpoint the contribution of the magnetic field intensity as the main driver of the FD.

\section{Discussions and Conclusion} 
\label{sec:conclusion}

In the present work, we study more than 300 ICME events that were detected over a period spanning over 17 years of ACE spacecraft data. A careful selection was made to remove: 1) low quality data, 2) interacting structures, 3) ICMEs non related to a clear FD (the latter formed a group that we used in Section \ref{sec:fd_nofd}). The FD data were taken from two neutron monitors at McMurdo and Oulu stations, i.e. in the two hemispheres but with similar rigidity cut-offs, for comparison. This allows us to ensure that the evolutions seen in the neutron monitor counts are not due to any instrumental specificities that could skew the results.
Finally, the data were prepared so as to only conserve isolated FDs (so as to not influence the background GCR level) and a careful removal of all GLEs.

We first performed a SEA on ICMEs by investigating the differences in the effects of the sheath, versus the ME. In particular, we showed that isolated ME (i.e., ICMEs without a sheath region) can be as efficient at driving a FD as a sheath. In the majority of cases however, ICMEs are associated with a sheath, which presence overshadows the impact of the FD driven by the ME alone. By comparing ME of similar mean speeds, with or without a sheath, we found that the speed profiles were also similar {(a consequence of ICME expansion physics). This allowed us to remove} the effect of the speed on the FD profile. 

Then, we conclude that the level of magnetic fluctuations (as indicated by the level of $rmsB${, which is generally observed to enhance in the sheath region}) is not a necessary condition to drive a FD, however, the intensity of the magnetic field{, whether associated with the sheath or the ME,} is. Furthermore, MEs without a sheath drive FDs that have a symmetrical profile (similar to that of the ME magnetic field), with the minimum flux close to the center of the structure (rather than at its front edge, as is found for MEs with a sheath), a picture consistent with that predicted by the model \citep{Dumbovic2018}. 
On the contrary, the presence of a sheath introduces a strong asymmetry in the GCR level in the following ME. This is a generalization of \cite{Belov2015}: their fast MCs with an asymmetric FD profile is due to the presence of a sheath at their front.

We also analyzed the recovery phase in the wake of the solar wind following ICMEs. We found that the recovery phase takes more than three times the entire ME length to go back to the pre-ICME levels. {This long recovery is also present behind ICMEs without a sheath, while at most a weak decrease of GCR flux is observed before these ICMEs.  These cases allow to exclude the sheath, the ICME expansion and the asymmetry of the magnetic field profile as the main sources of FD asymmetry before and after ICMEs.}  We conclude that the lower GCR level attained in the wake is mainly due to the spatial asymmetry of the GCR flux {background} since the main contribution here is the GCRs that arrive from behind the ME. This suggests that at 1 au, we are therefore able to witness the gradient of the GCR flux with FDs, {in agreement with what} was inferred from multi-spacecraft measurements in the literature.

We also extended the study to compare ICMEs with and without a sheath, driving or not a FD, so as to pinpoint which parameter could be responsible for driving the FD. Again, the classification by choosing the nearest ME speed in samples of ICMEs associated or not with a FD showed that the speed profile is similar, so this procedure allows us to well remove  the speed contribution to the FD (contrary to samples for which the selection is made with $<B_{ME}>$, see Appendix \ref{sec:appendix}).  Comparing plasma and magnetic profiles, we found that none of the plasma parameters, nor the rmsB, can explain the profiles of the FD.  With comparable velocity profiles, a FD is in fact associated with a high magnetic field intensity. In other words, we suggest that ICMEs with or without a sheath that have a low magnetic field intensity do not {typically} drive a FD, regardless of the rmsB and plasma profiles/intensities. 
Such a study therefore allows us to point to the essential ingredient in the drive of FDs, namely the velocity and the magnetic field intensity/profile. Moreover, the different substructures of an ICME (sheath and ME) can both drive FDs to equivalent levels. 

Our results confirm and observationally complement that of \cite{Benella2020}, where the authors used a Grad-Shafranov reconstruction for a magnetic cloud investigated in an event detected by the Wind spacecraft in August 2016. By using a full-orbit test particle simulation with this reconstruction, they were able to compute the trajectory of high energy particles. They found that small-scale magnetic fluctuations do not play a major role in the cross-field transport inside the magnetic structure, while it is most likely that that only the global magnetic configuration is needed to explain the observed FD.

A more thorough investigation will however be needed in the future to assess what local mechanisms are at play in both sub-structures. As was discussed in \citet{Wibberenz1998}, different processes can be put forward in the sheath and the ME. Most probably, since the diffusion coefficient is proportional to 1/$B$, diffusion reduces with an enhanced magnetic field as we find in both sheaths and MEs. 
Furthermore, new studies assessing the level of magnetic fluctuations in ICMEs \citep[such as][]{Kilpua2021} will help theoretically assess which part of the magnetic variation spectrum could potentially play a role in affecting the trajectory of GCR within ICME substructures.
As pointed out in \cite{Benella2020}, the curvature of the magnetic field may also play an important role in the GCR shielding effect. In addition, the type of magnetic connectivity of ICME substructures, i.e. close versus open, depending on the interaction of the magnetic field interaction with the surrounding magnetic field (e.g. erosion in the back or the front region of a ME, \cite{Ruffenach2015}, is expected to play a key role in the ability of GCRs to enter within MEs.

Finally, in the future, the increase of ICME detections at different heliodistances, for example closer to the Sun with the Parker Solar Probe \citep{Fox2016}, Solar Orbiter \citep{Muller2020} and Bepi-Colombo \citep{Benkhoff2010} missions, will give an exciting view on the relation between the magnetic field and the GCR interactions closer to the Sun. The present study also provides a roadmap to revisit previous results in the light of these new findings.

\acknowledgments
{We thank the referee for helpful suggestions which improve the clarity of the manuscript.} 
M.J. and F.R. acknowledge support from the Paris Domaine d’Int\'er\^et Majeur en Astrophysique et Conditions d’Apparition de la Vie+ (DIM ACAV+) financial support for the SOLHELIO project. M.J. and S.T-M. acknowledge support from the Erasmus+ 2021 traineeship program. 
J.G thanks the Strategic Priority Program of the Chinese Academy of Sciences (Grant No. XDB41000000), and the National Natural Science Foundation of China (Grant No. 42074222). {S.D. and C.G. acknowledge support from the Argentinean grants PICT-2019-02754 (FONCyT-ANPCyT), 
and UBACyT-20020190100247BA (UBA).}
We acknowledge the NMDB database (www.nmdb.eu), founded under the European Union's FP7 programme (contract no. 213007) for providing data for both McMurdo and Oulu neutron monitors.
We recognize the collaborative and open nature of knowledge creation and dissemination, under the control of the academic community as expressed by Camille No\^us at http://www.cogitamus.fr/indexen.html. The Cogitamus laboratory is a delocalized institution, bringing together scientists from all disciplinary backgrounds and nationalities who share common values: these of a disinterested research, scrupulous of integrity, aspiring to create, perpetuate, revise and transmit knowledge.

\vspace{5mm}
\facilities{McMurdo, Oulu, neutron monitors, NASA/ACE}

\software{astropy \citep{Astropy2013},  
          }

\appendix

\section{Comparison for ICMEs driving or not a FD with similar $B$}
\label{sec:appendix}

\begin{figure}[ht!]             
\plotone{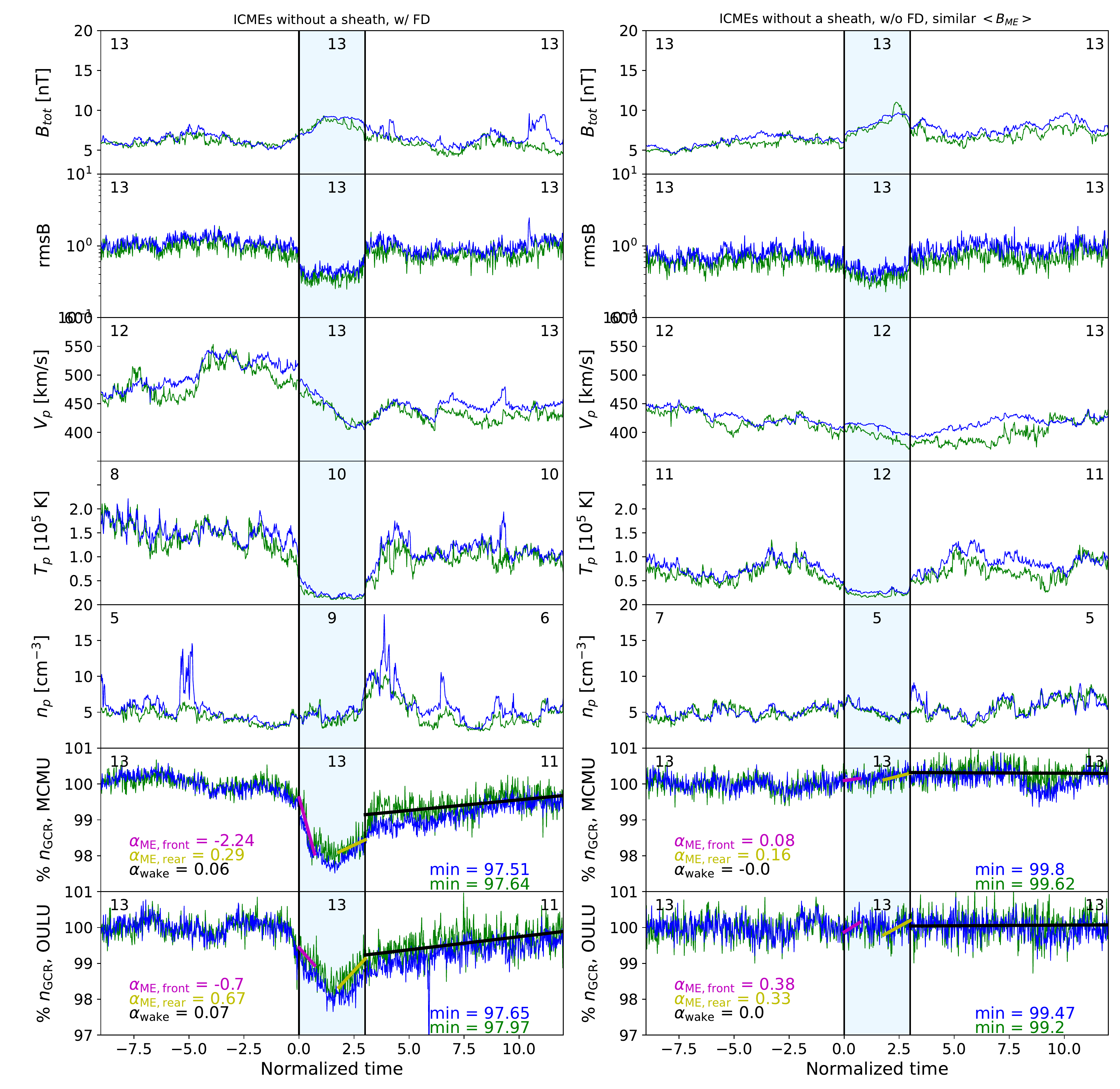}
\caption{Superposed epoch analysis for ICMEs without a sheath. (left) 13 ICMEs with a FD. (right) 13 other ICMEs without a FD and that have the closest ME mean magnetic field intensity from the ones on the left side. 
The parameters and the drawing conventions are the same as in Figure \ref{fig:icme_ws_ns_fd}.}
\label{fig:icme_ns_fdnofd_mag}
\end{figure}

\begin{figure}[t!]           
\plotone{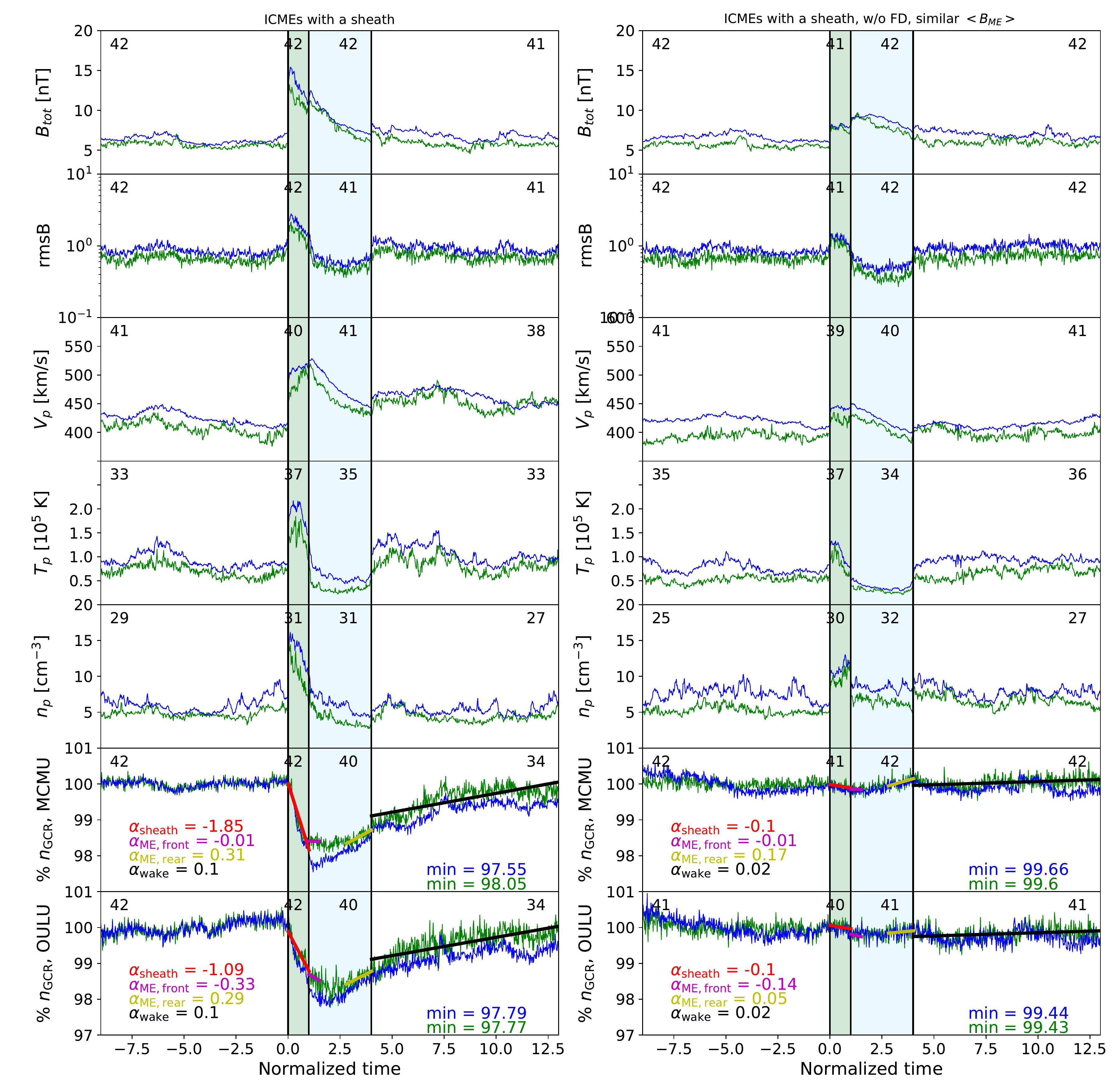}
\caption{Superposed epoch analysis for ICMEs with a sheath. (left) 42 ICMEs with a FD. (right) 42 other ICMEs without a FD and that have the closest ME mean magnetic field intensity from the ones on the left side.
The parameters {and the drawing conventions} are the same as in Figure \ref{fig:icme_ws_ns_fd}.}
\label{fig:icme_ws_fdnofd_mag}
\end{figure}

Figure \ref{fig:icme_ns_fdnofd_mag} shows in a similar fashion as Figure \ref{fig:icme_ns_fdnofd_speed} the samples of ICMEs with no sheath driving (left) or not (right) a FD, with similar $<B_{ME}>$ in the ME, corresponding to the magnetic field amplitude averaged within the ME.  We remark that the two samples, while having the same averaged $B$ intensity, do not have the same magnetic field profile in the ME: the magnetic field profile is more symmetrical for ICMEs associated with a FD, compared to the asymmetric, monotonously increasing profile found for ICMEs not associated with a FD. Next, the plasma parameters as well as the rmsB have similar values, except from the ME speed, which is larger for MEs driving a FD. Then, by selecting cases with the same averaged $B$ intensity, we outline here the role of the velocity. 
ICMEs which travel through a very fast solar wind as in the left panel had a short transit time towards 1 AU and their associated magnetic ejecta is still relatively ``empty'' with GCRs. On the right panel, with a longer transit time, the flux rope is already ``full" with GCRs, i.e., with the same level of GCR density as the ambient solar wind. Since the magnetic field and speed profiles are both different in the two samples, the separation of both contributions is a little less clear than with Figure \ref{fig:icme_ns_fdnofd_speed}. 

We also checked two samples of ICMEs
{, both of them containing sheath,} driving or not a FD, this time by selecting {cases with} the closest $<B_{ME}>$ (Figure \ref{fig:icme_ws_fdnofd_mag}). Similarly as what was concluded above for the ICMEs with no sheath, this shows the importance of velocity in driving a FD. 
The main differences in both profiles of $B$, as well as the intensities and profiles of $V_p$ in both the sheath and the ME mixes {the contributions of these two parameters. This limits our conclusion} compared to the results of Figure \ref{fig:icme_ws_fdnofd_speed}.

\bibliography{ICMEGCR}{}
\bibliographystyle{aasjournal}

\end{document}